\documentclass[aip,cha,reprint,runinaddress]{revtex4-1}
\usepackage{graphicx}
\usepackage{pifont}
\usepackage{amssymb}
\usepackage{xspace}
\usepackage[nativepdf,colorlinks=true,citecolor=blue,linkcolor=blue,anchorcolor=blue,urlcolor=blue]{hyperref}
\usepackage[absolute]{textpos}
\usepackage{amsfonts}

\begin{document}

\newcommand{\Ntrue}{$\mathcal{N}$\xspace}
\newcommand{\Nmeas}{$\mathcal{N}^*$\xspace}

\title{From brain to earth and climate systems: Small-world interaction networks or not?}
\author{Stephan Bialonski}
\email{bialonski@gmx.net}
\author{Marie-Therese Horstmann}
\author{Klaus Lehnertz}
\email{klaus.lehnertz@ukb.uni-bonn.de}
\affiliation{Department of Epileptology, University of Bonn, Sigmund-Freud-Str. 25, 53105 Bonn,}
\affiliation{Helmholtz Institute for Radiation and Nuclear Physics, University of Bonn, Nu\ss allee 14-16, 53115 Bonn,}
\affiliation{Interdisciplinary Center for Complex Systems, University of Bonn, R\"omerstr. 164, 53117 Bonn, Germany }

\received{22 December 2009}
\accepted{22 February 2010}
\published{31 March 2010}
\begin{abstract}
We consider recent reports on small-world topologies of interaction networks derived from the dynamics of spatially extended systems that are investigated in diverse scientific fields such as neurosciences, geophysics, or meteorology. With numerical simulations that mimic typical experimental situations we have identified an important constraint when characterizing such networks: indications of a small-world topology can be expected solely due to the spatial sampling of the system along with commonly used time series analysis based approaches to network characterization.
\end{abstract}
\pacs{05.90.+m, 02.40.Pc, 89.75.-k, 05.45.Xt, 05.45.Tp}
\maketitle

\begin{textblock*}{15cm}(3cm,26cm)
Copyright (2010) American Institute of Physics. This article may be downloaded for personal use only. Any other use requires prior permission of the author and the American Institute of Physics. The article appeared in Chaos 20, 013134 (2010) and may be found at \url{http://link.aip.org/link/doi/10.1063/1.3360561} --- DOI: 10.1063/1.3360561
\end{textblock*}

\begin{quotation}
The complex dynamics of spatially extended natural systems such as the human brain or the climate and the earth system is notoriously difficult to understand and has repeatedly stimulated a variety of scientific efforts, among them the development of sophisticated data analysis methods. Concepts from network theory promise to improve our understanding of such complex systems. These approaches consider a system as being composed of dynamically interacting subsystems whose functional interdependencies are reflected as links in an \emph{interaction network}. Interaction networks derived from field data of different dynamical systems have been consistently reported to possess small-world characteristics --- high local clustering and small average shortest path length --- which is considered as an indication of a common organization principle of natural dynamical systems. However, little attention has so far been paid to the conditions and assumptions underlying such analysis approaches. We here demonstrate that indications of small-world characteristics of interaction networks can solely be expected due to the spatial sampling of dynamical systems together with commonly used time series analysis techniques for the characterization of network links. Our findings not only call for the development and use of methods taking into account the spatial sampling of the studied systems but also for a careful interpretation and re-consideration of analysis results obtained so far.
\end{quotation}

\section{Introduction}

Over the last decade, network theory has contributed significantly to improve our understanding of complex systems, with wide applications in diverse fields, ranging from physics to biology and medicine \cite{Strogatz2001,Albert2002,Newman2003,Boccaletti2006a,Arenas2008}. 
With the introduction of the small-world network model by Watts and Strogatz \cite{Watts1998} research into complex networks has gained strong momentum. Since then, numerous studies have shown that a wide range of real world networks can be regarded as small-world networks (SWN) that are characterized by high levels of local clustering among nodes and by short paths that globally link nodes. Global and local network properties can be assessed with different measures \cite{Boccaletti2006a} among which the average shortest path length $L$ and the clustering coefficient $C$ are widely used in field studies. $L$ is the average shortest distance between any pair of nodes, and $C$ characterizes the local interconnectedness of nodes \cite{Watts1998}. Large values of both $L$ and $C$ are typical for lattices while low values of $L$ and $C$ are found for random networks. A low value of $L$ (or, from a network theoretic point of view, $L$ scaling at most logarithmically with the number of nodes $N$) and a high value of $C$ is usually considered as indicative for SWN\cite{Boccaletti2006a}.

Representing a complex system as a network requires identification of nodes and links which can be achieved straightforwardly in many scientific fields.  However, when investigating the dynamics of spatially extended systems in terms of complex networks -- such as in the neurosciences, in geophysics, or in meteorology -- identification of nodes and links is a challenging issue. Nodes are usually assumed to represent distinct subsystems and links represent interactions between them, and these nodes and links constitute an \emph{interaction network}. Lacking explicit knowledge of the structural organization of the dynamical system, nodes are usually associated with sensors that are placed so as to sufficiently capture the subsystems' dynamics, thereby considering theorems for an appropriate spatial and temporal sampling. While the latter is usually not an issue, choosing the right number of sensors and placing them in a meaningful way is highly nontrivial. When characterizing links, one is often faced with the problem that the underlying equations of motion are not known or that interactions between subsystems can not directly be measured experimentally. In these cases, time series analysis techniques are employed that aim at quantifying linear or nonlinear interdependencies between observables of subsystems recorded at the sensors \cite{Brillinger1981,Pikovsky_Book2001,Boccaletti2002,Hlavackova2007}. Eventually, weighted or binary (e.g., via thresholding), mostly undirected networks are constructed, and inference about small-world characteristics is usually based on a comparison of experimentally derived values of $L$ and $C$ to those of corresponding randomized networks. Using this approach small-world characteristics have been repeatedly reported for brain functional networks under both physiologic and pathophysiologic conditions \cite{Reijneveld2007,Bullmore2009}, and for seismic \cite{Abe2006,Jimenez2008} and climate networks \cite{Tsonis2004,Tsonis2008b,Donges2009}. Despite the use of highly sophisticated analysis techniques, the aforementioned challenging issue of properly identifying nodes and links of interaction networks remains largely unsolved.

We here demonstrate --  by simulating typical experimental situations in the diverse scientific fields --
that indications of a small-world topology can be expected solely due to the spatial sampling together with influencing factors that can be attributed to the widely used data-driven approaches to link and network characterization.

\section{Inferring and characterizing interaction networks from field data}
We begin with recalling approaches that have been employed very frequently to derive and analyze interaction networks. First, interdependencies between subsystems are estimated using some bivariate time series analysis technique. Among the many available methods\cite{Brillinger1981,Pikovsky_Book2001,Boccaletti2002,Kantz2003,Pereda2005,Hlavackova2007,Lehnertz2009b} 
particularly symmetric techniques that aim at assessing the strength of interaction $m_{ij}$ between time series $i$ and $j$ are often used to derive undirected binary networks.

Second, from matrix $\mathbf{M}$ with entries $m_{ij}=m_{ji}$ the adjacency matrix $\mathbf{A}$ representing the assumed underlying interaction network is derived by thresholding. Let $N$ denote the number of nodes of the interaction network and let $k_i$ denote the degree of node $i$, i.e., the number of nodes to which node $i$ is connected. In most field studies, the mean degree $k=N^{-1}\sum_i k_i$ is chosen and entries $a_{ij}$ of $\mathbf{A}$ are set to 1 for all $kN$ largest entries $m_{ij}$ (resulting in $kN/2$ undirected links), and $a_{ij}=0$ (no link) otherwise.

Third, clustering coefficient $C$ and average shortest path length $L$ are determined to characterize the interaction network and to assess its possible small-world characteristics. The local clustering coefficient $C_i$ quantifies the local interconnectedness of the network and is defined as the fraction of the number of existing links between neighbors of node $i$ among all possible links between these neighbors\cite{Watts1998,Newman2003,Boccaletti2006a}
\begin{equation}
 C_i = \left\{ \begin{array}{cl} \frac{1}{k_i (k_i-1)} \sum_{j,m} a_{ij} a_{jm} a_{mi}, & \mbox{if }k_i > 1\\ 0, & \mbox{if } k_i \in \{0,1\}\mbox{.}\end{array}\right.
\end{equation} 
The clustering coefficient $C$ of the network is then defined as the mean of the local clustering coefficients
\begin{equation}
 C = \frac{1}{N} \sum_{i=1}^N C_i
\end{equation}
and represents a local scale property of the network. Note that $C_i,C \in [0,1]$.

The average shortest path length represents a global scale property of a network and is defined as the average distance between any two nodes,
\begin{equation}\label{eq:L1}
 \hat{L} = \frac{2}{N(N+1)} \sum_{i\leq j} l_{ij}\mbox{,}
\end{equation}
where $l_{ij}$ denotes the length of the shortest path between node $i$ and $j$. Note that some authors include the distance from each node to itself in the average (as we do here, $l_{ii} = 0$), while others exclude it\cite{Newman2003}. Exclusion will, however, only change the value of $\hat{L}$ by the constant factor of $(N+1)/(N-1)$. An issue which may be encountered when determining $\hat{L}$ for interaction networks derived from field data are pairs of nodes which do not possess a connecting path, in which case $l_{ij} = \infty$. To avoid this problem, it has been proposed to replace $l_{ij}$ in Eq. \ref{eq:L1} with $l_{ij}^{-1}$ resulting in a quantity called efficiency\cite{Latora2001,Latora2003}. Another approach, which is often used in field studies, is to exclude infinite $l_{ij}$ from the average in Eq. \ref{eq:L1}. We here follow this ansatz and define the average shortest path length $L$,
\begin{equation}
 L = \frac{1}{N_l} \sum_{(i,j) \in S} l_{ij}\mbox{,}
\end{equation}
where 
\begin{equation}
 S = \{(i,j) \mid l_{ij} < \infty;\mbox{ } i,j = 1,\ldots,N \}
\end{equation}
denotes the set of all pairs of nodes $(i,j)$ which are connected through some path with finite path length $l_{ij}$, and where $N_l = \mid S \mid$ denote the number of such pairs.

From a network theoretic point of view\cite{Watts1998,Watts1999}, $L$ scaling at most logarithmically with $N$, and a high value of $C$ is considered the hallmark of SWN topology. This approach, however, comes along with a requirement that can hardly be fulfilled in typical experimental situations.
In some cases it is impossible to vary the number of sensors (i.e. number of nodes) due to experimental restrictions, while in other cases it might be the nature of the system itself that does not allow the recording of observables from an arbitrary number of sensors. Moreover, even if it were possible to vary the number of sensors, this would require choosing which sensor to remove or where to add an additional sensor --- a choice that is highly dependent on a priori knowledge of the spatial organization of the dynamical system. It is thus not astonishing, that field studies do not evaluate a possible scaling behavior of $L$. Instead, in the fourth step values of $C$ and $L$ are typically compared with values $\langle C_r\rangle$ and $\langle L_r\rangle$ obtained from ensembles of the corresponding random networks \cite{Maslov2002,Maslov2004} that are generated by randomizing the original network while preserving the degree of each node. Eventually, values of $\gamma := C/\langle C_{\rm{r}} \rangle > 1$ and $\lambda := L/ \langle L_{\rm{r}} \rangle \approx 1$ are assumed to be indicative of SWN. 

\subsection*{An illustrative example from field data analysis}
\begin{figure}
\includegraphics{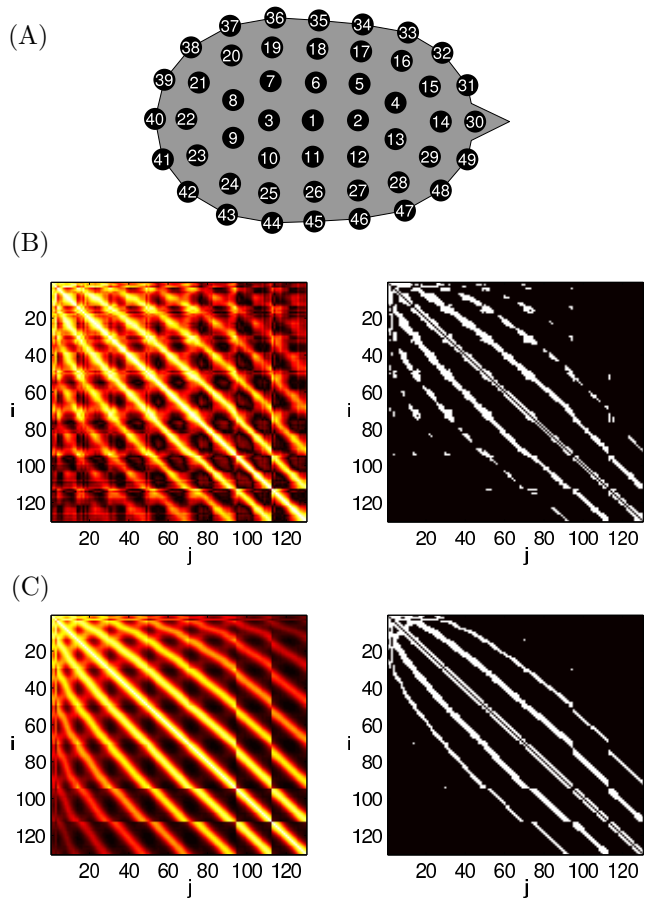}
\caption{\label{fig1} (Color online) (A) Schematic of the spatial arrangement of a subset of sensors used to sample the dynamics of a human brain by MEG. (B) Left: Exemplary matrix $\mathbf{M}$ where entry $m_{ij}=m_{ji}$ is the absolute value of the correlation coefficient between MEG time series $x_i(t)$ and $x_j(t)$ from sensor pair $(i,j)$. Right: Adjacency matrix $\mathbf{A}$ derived from $\mathbf{M}$ by thresholding with $k=15$. (C) Left: Matrix $\mathbf{\tilde{M}}$ with entries $\tilde{m}_{ij}=F(d_{ij})$, where $d_{ij}$ denotes the Euclidean distance between sensors $i$ and $j$ in 3-dimensional space, and $F(d_{ij}) = (1+\exp(u(d_{ij}-v)))^{-1}$ with $u=23$ and $v=0.1$. Right: $\mathbf{\tilde{A}}$ derived from $\mathbf{\tilde{M}}$ by thresholding with $k=15$. Note that $\mathbf{\tilde{A}}$ is not affected by the choice of $F$, as long as $F$ decreases strictly monotonically with increasing $d_{ij}$. Entries of all matrices range from 0 (black) to 1 (white).}
\end{figure}

Using the aforementioned approaches for inferring and characterizing interaction networks we analyzed time series of brain magnetic activities (magnetoencephalography; MEG \cite{Hamalainen1993}) that were recorded (sampling rate: 254.31 Hz; 16 bit A/D conversion; bandwidth: 0.1--50 Hz; 148-channel magnetometer system, cf. Fig. \ref{fig1}A)) from a healthy subject with eyes closed\cite{Horstmann2010} (the subject had signed informed consent that the data might be used and published for research purposes, and the study was approved by the local medical ethics committee).
Here we restrict ourselves to $N=130$ sensors in order to minimize contaminations with muscle activity in the lowermost ring of sensors. As an interdependence measure, here we choose the absolute value of the correlation coefficient,
\begin{equation}
 \rho_{ij} = T^{-1} \left| \sum_t (x_i(t)-\langle{x}_i\rangle)(x_j(t)-\langle{x}_j\rangle)\sigma_i^{-1}\sigma_j^{-1} \right| \mbox{,} 
\end{equation}
where $x_i(t)$ and $x_j(t)$ denote MEG time series of length $T$ ($T=4096$ data points) recorded at sensors $i$ and $j$, respectively, and $\langle{x}\rangle$ and $\sigma$ denote the respective mean values and standard deviations. In Fig. \ref{fig1}B we show the matrix $\mathbf{M}$ with entries $m_{ij}=\rho_{ij}$.
We derive the adjacency matrix $\mathbf{A}$ (with entries $a_{ij}$) of the interaction network by choosing a mean degree $k$ and thresholding as described above. $\mathbf{A}$ displays a pattern of diagonals, which can also be found in $\mathbf{M}$ and can be attributed to spatially close sensors (cf. Fig. \ref{fig1}A). For $\mathbf{A}$ we obtain $C=0.58$ and $L=3.13$. As in many field studies, we proceed by comparing values of $C$ and $L$ with those of 100 realizations of corresponding random networks \cite{Maslov2002,Maslov2004} and assume $\gamma > 1$ and $\lambda \approx 1$ to be indicative of SWN. With $\gamma = 4.21 \pm 0.15$ and $\lambda = 1.53 \pm 0.01$ this interaction network would have been interpreted as SWN.

Let us now consider some interdependence measure $\tilde{m}$ that depends on the Euclidean distance in 3-dimensional space $d_{ij}$ between sensors only. We assume $\tilde{m}$ to decrease strictly monotonically with $d_{ij}$ such that $\tilde{m}$ from spatially close sensors will attain larger values than $\tilde{m}$ from spatially distant sensors. This leads to a network with distance-dependent connectivity structure, an example of a spatial network\cite{Boccaletti2006a,Costa2007}. In Fig. \ref{fig1}C we show that both $\mathbf{\tilde{M}}$ and $\mathbf{\tilde{A}}$ display characteristic diagonal patterns already observed in the respective matrices derived from field data. For $\mathbf{\tilde{A}}$ we obtain $C=0.57$ and $L=3.14$, and with $\gamma = 4.97 \pm 0.18$ and $\lambda = 1.55 \pm 0.01$ even this network would have been interpreted as SWN.

\section{Investigation of influencing factors}
There are a number of influencing factors that can adversely affect the assessment of small-world characteristics of interaction networks derived from field data. In addition to the unavoidable imprecision of the data acquisition, the accuracy of some interdependence measure used to characterize links is usually restricted due to the limited amount of accessible data and is spoiled due to unavoidable noise contributions. Together with thresholding methods for deriving interaction networks --- for which the mean degree is often chosen empirically --- this may lead to spurious additional links present in the network as well as to spuriously missing links. Thus, the question arises as to how reliable do we have to estimate links in order to safely infer small-world characteristics of interaction networks from field data? Moreover, strongly interdependent signals may either reflect some functional interaction between different subsystems or may be caused by sampling the same subsystem, but most time series analysis techniques do not allow one to unequivocally distinguish between both cases. How does this affect the inference of small-world characteristics of interaction networks, even in cases where such a distinction was in principle possible? In the following paragraphs we address these questions in more detail.

\begin{figure}[tb]
\includegraphics[width=240pt]{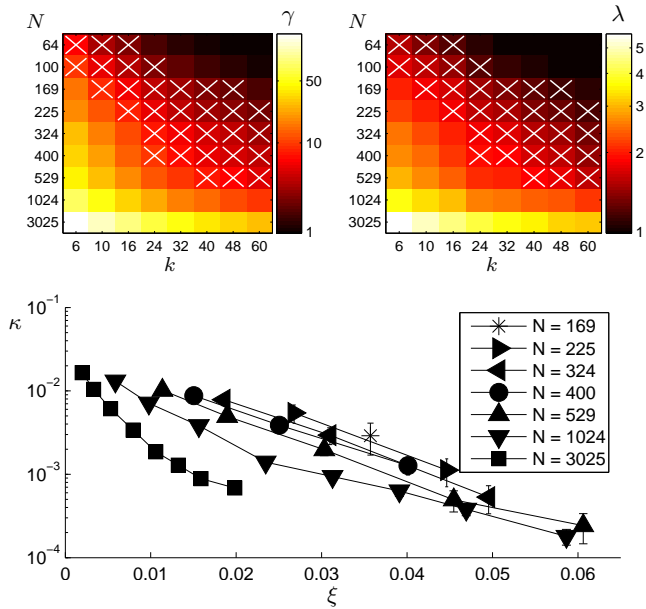}
\caption{\label{fig2} (Color online)
Top: Mean values of normalized clustering coefficient $\gamma$ (left) and normalized average shortest path length $\lambda$ (right) for two-dimensional square lattices with different numbers of nodes $N$ and mean degrees $k$ (maximum standard deviations: $\sigma_{\gamma}=0.02$ and $\sigma_{\lambda}=0.02$). White crosses mark $(N,k)$ configurations for which lattices will be classified as SWN if $\gamma > 2$ and $\lambda < 2$ is chosen as a practical criterion. Bottom: Minimum fraction of randomly replaced links $\kappa$ for which the resulting network would be classified as SWN ($\lambda < 2$) in dependence on the density of links $\xi$. Error bars denote standard deviations derived from 100 independent replacement runs, and lines are for eye guidance only. Note that error bars are smaller than symbol size in the majority of cases.}
\end{figure}

\subsection*{Impact of measurement uncertainties and latticelike arrangement of sensors}

We assume $N$ sensors to be arranged on a square-lattice -- which is quite typical for field studies -- and associate sensors with nodes. Taking into account that interactions between spatially neighbored nodes are usually stronger than spatially distant nodes (spatial correlations), we assume an interdependence measure to decrease strictly monotonically with an increasing Euclidean distance between sensors only. We derive links via thresholding with mean degree $k$. Note, that the derived networks possess lattice topologies due to the construction scheme. We chose values of $N$ and $k$ as typically reported in field studies and estimate $\gamma$ and $\lambda$ for the derived networks as above. For a range of $(N,k)$ values (cf. Fig. \ref{fig2}, upper part), we observe $\gamma \gg 1$ and $\lambda \approx 1$, which would indicate these networks to possess small-world characteristics. For parameters from the upper right region of the $(N,k)$ plane, networks would not be classified as SWN since $\gamma$ approaches $1$ with an increasing density of links $\xi := k/(N-1)$. For sparse networks (lower left region of the $(N,k)$ plane) we observe $\lambda \gg 1$ for a range of $(N,k)$ values, which also would not indicate SWN. For these sparse networks, however, the reliability of estimating links is of crucial importance for a correct classification.

In principle, a limited reliability of link estimation will lead to spurious additional links and to spurious missing links contributing to the topology of the network. Controlling the amount of spurious links could be achieved by multiple testing against some appropriately chosen null model\cite{Kramer2009,Donges2009b}. Such an approach allows to control the probability of erroneously detecting links (false positives), but it is well known for its limited power which leads to a starkly increased number of false negatives (missing links). In addition, taking into account problems of defining appropriate null models for time series (i.e. surrogates\cite{Schreiber2000a}) and the computational burden of generating them, multiple testing procedures have not found wide applicability in studies of interaction networks derived from field data. Nevertheless, we can carry over concepts from multiple testing methods to learn about the reliability needed to correctly infer from $\gamma$ and $\lambda$ that the networks in the lower left corner of the $(N,k)$ parameter space (cf. Fig. \ref{fig2} upper part) possess lattice topologies. For these sparse networks we model uncertainties from estimating links (i.e., the number of false positives) by randomly replacing links in the networks. With $n_{\rm r}$ we denote the average minimum number of randomly replaced links that would lead to a classification of the original lattice as SWN due to a decrease of the average shortest path length such that $1 \approx \lambda < \Delta$. As a practical criterion we here exemplarily set $\Delta = 2$. In the lower part of Fig. \ref{fig2} we show that a minimum fraction of randomly replaced links $\kappa := 2n_{\rm r}/(kN)$ of less than 2 \% would suffice to falsely classify these sparse networks as SWN (note that $\kappa$ represents the false discovery rate\cite{Benjamini1995} in the context of multiple testing procedures). The replacement of links affects $\gamma$ only marginally, and we always observed $\gamma \gg 1$. Note that $\kappa$ even decreases with an increasing density of links $\xi$. Particularly for networks with a small number of nodes and depending on the chosen mean degree we observe that only one to five randomly replaced links lead to $\lambda < \Delta$. We note that together with theoretical arguments\cite{Newman1999,Barthelemy1999,*Barthelemy1999e,Petermann2006} our findings are decisive for a characterization of networks derived from field data. 
By construction, the average shortest path length $L$ depends sensitively on the actual link structure and changing or adding only a few links can result in a remarkable change of $L$. 

Summarizing, under unfavorable conditions the often used latticelike arrangement of sensors (we note that we obtained similar findings for three-dimensional lattices) together with a limited reliability of estimating links leads to indications of a small-world topology of interaction networks derived from the dynamics of spatially extended systems, even if the underlying interaction structure is not SWN.

\begin{figure}[tb]
 \includegraphics{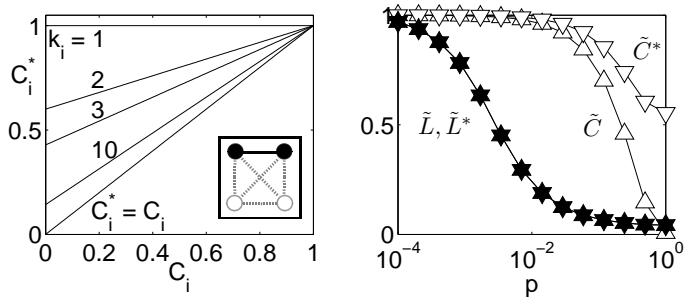}
\caption{\label{fig3} Left: Local clustering coefficient $C_i^*$ of node $i$ of \Nmeas as a function of $C_i$ of \Ntrue for different node degrees $k_i$. Construction of \Nmeas is shown schematically in the inset. Nodes and links included in \Ntrue and \Nmeas are colored black, while nodes and links included in \Nmeas only are colored gray. Right: Means of $\tilde{C}(p) := C(p)/C(0)$ (open symbols) and $\tilde{L}(p) := L(p)/L(0)$ (filled symbols) for \Ntrue depending on the rewiring probability $p$ (lines are for eye guidance only). $\tilde{C}^*(p)$ and $\tilde{L}^*(p)$ denote the corresponding quantities for \Nmeas. We used the Watts-Strogatz scheme ($N=1000$, $k=4$, 1000 realizations for each $p$) to generate \Ntrue networks (symbol $\bigtriangleup$) and derived \Nmeas networks (symbol $\bigtriangledown$) by duplicating all nodes from \Ntrue. Standard deviations for all quantities are smaller than symbol size.}
\end{figure}

\subsection*{Impact of common sources}
We now consider time series analysis techniques which can not distinguish between interdependencies due to functional interactions between different subsystems and interdependencies caused by sampling the same subsystem (i.e., a common source). We address the question of how these techniques affect the inference of small-world characteristics. 
We assume that a given system can be regarded as some network \Ntrue with some topology and that \Ntrue consists of $N$ nodes and some number of links. Ideally, one would sample the system by placing $N$ sensors such that each captures the dynamics of the respective subsystem. In field studies, however, the number of subsystems reflecting the dynamical organization of \Ntrue is usually not known a priori and we would choose the number of sensors such that they allow a reasonably dense spatial sampling. This might come along with the risk of sampling the dynamics of the same subsystem by two or more sensors.
With the time series analysis techniques we have in mind (such as cross correlation, coherency, mean phase coherence, mutual information), this will lead to an indication of strongly interdependent time series. We simulate this situation by introducing for each ideally placed sensor $i$ ($i \in {1,\ldots,N}$) another sensor $i'$ with zero spatial distance between $i$ and $i'$. This leads to a network \Nmeas possessing $N^*=2N$ nodes. Each duplicate node $i'$ inherits the same neighbors of $i$ and is connected to $i$ (cf. inset in left part of Fig. \ref{fig3}). We derive $C_i^*$ and $L^*$ of \Nmeas as functions of $C_i$ and $L$ of \Ntrue as

\begin{eqnarray}
C_i^* &=& \left\{\begin{array}{cl} \frac{3}{2k_i+1}+2C_i\frac{k_i-1}{2k_i+1}, & \mbox{if }k_i > 0\\ 0, & \mbox{if } k_i=0\end{array}\right.\\
L^* &=& L + L_1^+ \mbox{ with } L_1^+ = \frac{N}{2N_l}\mbox{,} \label{eq:2}
\end{eqnarray}
where $k_i$ and $N_l$ are quantities of \Ntrue. Note that $L_1^+ \in [\frac{1}{2N}, \frac{1}{2}]$ where the lower bound holds for connected networks (a path exists between every pair of nodes) and the upper bound for networks without links. Obviously the impact of introducing additional nodes (i.e., sensors) on the average shortest path length can be neglected, since $L^* \approx L$. In contrast, the clustering coefficient $C^*=({N^*}^{-1}\sum_{i=1}^{N^*} C_i^*) \geq C$ increases, since $C_i^* \geq C_i$ (cf. Fig. \ref{fig3} left). In the right part of Fig. \ref{fig3} we demonstrate this effect for different network topologies of \Ntrue that we derived by employing the construction scheme proposed by Watts and Strogatz\cite{Watts1998}. For all rewiring probabilities $p$ we observe $L^*(p)/L^*(0) \approx L(p)/L(0)$, however, $C^*(p)/C^*(0)$ clearly exceeds $C(p)/C(0)$ with increasing $p$ such that even \Nmeas derived from random networks \Ntrue ($p=1$) would have been characterized as SWN.

\begin{figure}[tb]
\includegraphics{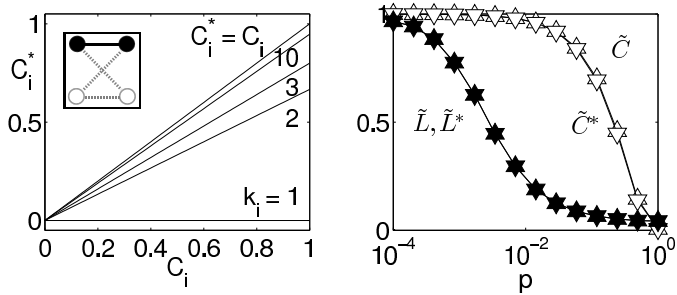}
\caption{\label{fig4} Left: Local clustering coefficient $C_i^*$ of node $i$ of \Nmeas as a function of $C_i$ of \Ntrue for different node degrees $k_i$. Construction of \Nmeas is shown schematically in the inset. Nodes and links included in \Ntrue and \Nmeas are colored black, while nodes and links included in \Nmeas only are colored gray. Right: Means of $\tilde{C}(p) := C(p)/C(0)$ (open symbols) and $\tilde{L}(p) := L(p)/L(0)$ (filled symbols) for \Ntrue depending on the rewiring probability $p$ (lines are for eye guidance only). $\tilde{C}^*(p)$ and $\tilde{L}^*(p)$ denote the corresponding quantities for \Nmeas. We used the Watts-Strogatz scheme ($N=1000$, $k=4$, 1000 realizations for each $p$) to generate \Ntrue networks (symbol $\bigtriangleup$) and derived \Nmeas networks (symbol $\bigtriangledown$) by duplicating all nodes from \Ntrue. Standard deviations for all quantities are smaller than symbol size.}
\end{figure}

We now evaluate the opposite situation and consider time series analysis techniques, which we assume to be able to distinguish between interdependencies due to functional interactions between different subsystems and interdependencies due to a common source. We proceed as above and introduce for each ideally placed sensor $i$ ($i \in {1,\ldots,N}$) another sensor $i'$ with zero spatial distance between $i$ and $i'$, which again leads to a network \Nmeas possessing $N^*=2N$ nodes. Each duplicate node $i'$ inherits the same neighbors of $i$ but --- different to the simulation before --- this duplicate node is not connected to $i$ (cf. inset in left part of Fig. \ref{fig4}). We derive $C_i^*$ and $L^*$ of \Nmeas as functions of $C_i$ and $L$ of \Ntrue as

\begin{eqnarray}
 C_i^* &=& C_i \frac{k_i-1}{k_i-\frac{1}{2}}  \mbox{,}\\
 L^* &=& l^* L  + L_2^+ \mbox{,}
\end{eqnarray}
where
\begin{equation}
 l^* = \left( 1 - \frac{N_0}{2N_l} \right)^{-1} \mbox{ and  } L_2^+ = \left( \frac{N-N_0}{N_l-\frac{1}{2}N_0} \right)\mbox{.}
\end{equation}
$N_0$ is the number of nodes in \Ntrue with no neighbors, i.e. $N_0 = \mid \{ i \mid k_i = 0, i = 1, \ldots, N\}\mid$. Note that $l^* \in [1,2]$ where the upper bound holds for the special case of networks \Ntrue without links ($N_0 = N$) and the lower bound for networks where each node possess at least one link ($N_0 = 0$), which, e.g., is the case for connected networks. $L_2^+ \in [0,\frac{1}{2}]$ where the lower bound holds for networks without links and is approached by connected networks ($L_2^+ = N^{-1}$) and the upper bound is approached by the special case of networks with decreasing $N_0$ and increasing number of connected components and reached for $N/2$ connected components and $N_0 = 0$. Again, the impact of introducing additional nodes (i.e., sensors) on the average shortest path length can be neglected in networks possessing links ($L^* \approx L$). The clustering coefficient $C^*$ decreases since $C_i^* \leq C_i$ depending on the corresponding degrees of the nodes (cf. Fig. \ref{fig4} left). 
Note, however, that the maximum possible reduction amounts to $C_i^*= \frac{2}{3}C_i$ ($k_i = 2$) only and that $C_i^* = C_i$ for $k_i \in \{0,1\}$ together with $C_i^* \rightarrow C_i$ for increasing $k_i$ will likely cause only a slight decrease of $C^*$ in real world networks.

In the right part of Fig. \ref{fig4} we present our findings for different network topologies of \Nmeas that we derived by employing the Watts-Strogatz construction scheme. For all rewiring probabilities $p$ we observe $L^*(p)/L^*(0) \approx L(p)/L(0)$ and $C^*(p)/C^*(0) \leq C(p)/C(0)$. Thus, \Nmeas derived from random networks \Ntrue ($p=1$) would not be characterized as SWN but as random networks when employing time series analysis techniques, which can unequivocally distinguish between interdependencies due to a common source and interdependencies due to functional interactions between different subsystems. 

To our knowledge, only few time series analysis techniques have been published that address the problem of interdependencies due to sampling the same (or nearly the same) dynamical subsystem\cite{Nolte2004,Stam2007c}. These techniques are based on the assumption that common components in both time series stemming from the same dynamical subsystem will lead to instantaneous interdependencies (with zero time lag). Separating these instantaneous interdependencies from those associated with a non-zero time lag could lead to techniques capturing interdependencies between different interacting dynamical subsystems only. We note, however, that these techniques have not yet found broad application in field studies.

\section{Discussion}
Several influencing factors may hamper the unequivocal inference of small-world characteristics in interaction networks derived from field data using well accepted time series analysis techniques together with network theoretic approaches. Taken together, these influences will most likely lead to findings of small-world characteristics in interaction networks irrespective of their actual organization. These influences may be associated with four aspects which we discuss in the following.

First, the commonly used comparison of values of $C$ and $L$ with those obtained for corresponding random networks 
comes along with the risk of classifying even an actual lattice as SWN (cf. Fig. \ref{fig2} top). The reason is, that such a comparison can only provide clues as to how much the topology of the network under study differs from that of a random network but not from that of a lattice. A comparison with lattices has been proposed\cite{Sporns2004} but has not yet been found wide application. The latter can be attributed to the fact, that when comparing with lattice topologies one has to decide upon the dimensionality and construction of such lattices, both representing non-trivial choices, which can decisively influence the result of such a comparison.

Second, the addition of, the change in, or the deletion of only a few links --- a likely consequence of a limited reliability of estimating links --- can significantly change the value of average shortest path length $L$. In networks possessing latticelike topologies (large $C$ and $L$ values), uncertainties in link estimation can likely cause $L$ to decrease leading to small-world characteristics of the network under study (cf. Fig. \ref{fig2} bottom). In the light of a limited reliability of time series analysis techniques used to estimate links, multiple testing procedures can help control the probability of false positives in networks derived from field data\cite{Kramer2009}. Whereas methods controlling the family-wise error (i.e., the probability of detecting false links among all possible pairs of nodes) have been developed over the years but are known to come along with a high risk of false negatives\cite{Tamhane1996} and thus missing links, methods controlling the false-discovery rate (i.e., the probability of false positives among all inferred links) appear to be promising approaches with lower risk of false negatives\cite{Benjamini1995,Benjamini2001,Kramer2009}. However, limiting the probability of erroneously adding, changing, or deleting just a few links calls for small probabilities of both, detecting false positives as well as missing false negatives, which represents a challenging task for currently available multiple testing methods. In addition, hypothesis testing involves defining appropriate null-models for time series, which again represents a challenging issue\cite{Schreiber2000a,Thiel2006}.

Third, the placement of sensors to capture the dynamics of an unknown spatially extended complex system and their representation as nodes of an interaction network already represents an interpretation of the data which comes along with various pre-assumptions, e.g. the system can be meaningfully decomposed into subsystems and their dynamics is captured by the sensors. Missing to capture the dynamics of only a few of the subsystems may lead to a dramatic change in $L$ as discussed above. In addition, capturing the dynamics of the same subsystem with two or more sensors together with commonly used time series analysis techniques to measure interactions
can lead to an increase of the clustering coefficient $C$ of the network (cf. Fig. \ref{fig3}). This may lead to the inference of small-world characteristics even for interaction networks actually possessing a random network topology. We have demonstrated (cf. Fig. \ref{fig4}) that time series analysis techniques that would be capable of unequivocally distinguishing between interdependencies due to common sources and interdependencies due to interacting dynamical subsystems can overcome the problem of an artificial increase of the clustering coefficient.
In this context we mention two time series analysis techniques\cite{Nolte2004,Stam2007c} which purport to be capable of distinguishing between such cases. It remains to be shown, however, whether these as well as other time series analysis techniques are capable of distinguishing between direct and indirect interactions, another potential mechanism for a spuriously increased clustering coefficient. An alternative approach toward tackling the problem of a spuriously increased clustering coefficient is to manually correct $C$ for the influence of spatially sampling the same dynamical aspect\cite{Tsonis2008a}. This approach, however, relies on a priori knowledge about the system which may not be generally available.
A decomposition of systems -- which actually represent spatial diffusion or field processes -- into subsystems may introduce spatial correlations in the network topology and leads to the question as to how much the derived interaction networks depend on the coarse graining of the dynamics. In this case, interaction networks may not describe properties inherent to the studied dynamics but properties that solely depend on the applied coarse graining scheme. The value of such a description may vary and will depend on the application and aim of the study.

Fourth, already in the first small-world model of Watts and Strogatz\cite{Watts1998} the wide regime of small-world networks is flanked by the special cases of random and lattice networks. Together with the aforementioned influencing factors, findings of small-world interaction networks in an absolute sense should be expected and are not surprising. In contrast, when being able to eliminate all influencing factors, findings of the special (and thus possibly rare) topologies of random or lattice interaction networks would be surprising. Besides deciding upon small-world characteristics in an absolute sense, more recent studies\cite{Schindler2008a,Horstmann2010,Tsonis2008c} aim at studying relative changes of $C$ and $L$ only. In view of the wide regime of small-world networks, such an approach appears more promising.

\section{Conclusion}
We have identified an important constraint when analyzing interaction networks derived from the dynamics of spatially extended systems, namely that solely the spatial sampling of the system together with a time series analysis based link characterization can lead to indications of a small-world topology. The arrangement of sensors commonly used in field studies along with the ansatz of considering a spatially extended dynamical system as a complex network of interacting subsystems imposes a spatial structure on the system, irrespective of its actual organization, which may also underlie spatial restrictions. Given these constraints, it would be surprising to not observe small-world indications for interaction networks derived from spatially extended dynamical systems. Whether the actual interaction structure of these systems is indeed small-world or not --- a question addressed in various studies in the neurosciences, and the climate and earth sciences --- cannot be unequivocally answered with currently employed analysis techniques. We therefore consider it advisable to avoid interpretations of findings of small-world topologies for such interaction networks in an absolute sense.

Here we here restricted our investigations to widely used network characteristics, the clustering coefficient $C$ and average distance between nodes $L$. It is conceivable though, that the abovementioned constraint will likely affect other network characteristics. To advance our understanding of the dynamics of spatially extended systems we consider the following directions of research as promising: (i) improving the determination of the actual structural organization of a dynamical system can help to advise the design of appropriate sensor placement strategies (such an approach is currently being pursued in the neurosciences\cite{Hagmann2008}); (ii) developing novel and refining existing time series and network analysis techniques\cite{Stam2007c,Nolte2008,Sporns2004,Serrano2009,Fortunato2010} together with computational network analyses\cite{Arenas2006,Timme2007,Gfeller2008} can help to unravel functional interactions from contributions that result from sampling the same subsystem.

\begin{acknowledgments}
We thank Alexander Rothkegel and Andreas Hense for useful comments. S.B. acknowledges support from the German National Academic Foundation. M.T.H. and K.L. acknowledge support from the Deutsche Forschungsgemeinschaft (Grant No. LE 660/4-1).
\end{acknowledgments}

\end{document}